\documentclass[12pt,showpacs,showkeys,superscriptaddress,aps]{revtex4}
\usepackage{amsmath,amsthm,amssymb}
\usepackage{epic, eepic}

\usepackage{graphicx} 

\usepackage{braket}
\usepackage{bm}

\usepackage{mathrsfs,fancyhdr,ascmac}
\usepackage{url}

\usepackage{color}
\usepackage{here}
\newcommand{\maprightu}[1]{%
\smash{\mathop{%
\hbox to 1cm{\rightarrowfill}}\limits^{#1}}}
\newcommand{\maprightd}[1]{%
\smash{\mathop{%
\hbox to 1cm{\rightarrowfill}}\limits_{#1}}}
\newcommand{\mapleftu}[1]{%
\smash{\mathop{%
\hbox to 1cm{\leftarrowfill}}\limits^{#1}}}
\newcommand{\mapleftd}[1]{%
\smash{\mathop{%
\hbox to 1cm{\leftarrowfill}}\limits_{#1}}}
\newcommand{\mapnerssss}[1]{%
\smash{\mathop{%
\hbox to 3cm{\nearrow}}\limits^{#1}}}
\begin{document}
\title{Components of Invariant Variety of Periodic Points\\
and \\
Fundamental Domains of Recurrence Equation
}
\author{Tsukasa YUMIBAYASHI}
\email[email : ]{t.yumibayashi@kiso.phys.se.tmu.ac.jp}
\affiliation{Department of Physics, Tokyo Metropolitan University,\\
Minamiohsawa 1-1, Hachiohji, Tokyo, 192-0397 Japan}
\begin{abstract}
In this paper, we discuss duality about components of invariant variety of periodic points(IVPP)\cite{IVPPa}\cite{IVPPb}\cite{IVPPbook} and fundamental domain of recurrence equation, and present an algorithm for the derivation of all components of IVPPs of any rational maps. It is based on the study of two examples of a 2 dimensional map and a 3 dimensional map. In particular, all components of IVPPs of the 2 dimensional example are completely determined by means of the cyclotomic polynomials\cite{cp}.  
\end{abstract}
\maketitle
\section{Introduction}
The invariant variety of periodic points, or the IVPP for short\cite{IVPPa}\cite{IVPPb}\cite{IVPPbook}, of a $d$ dimensional map $F : \mathbb{C}^d \rightarrow \mathbb{C}^d$, $F : \bm{x}^t \mapsto \bm{x}^{t+1}, \bm{x}^t, \bm{x}^{t+1} \in \mathbb{C}^d$ with $p$ invariants $\bm{r}: \mathbb{C}^d \rightarrow \mathbb{C}^p, \ \mathrm{s.t.} \ \bm{r}(\bm{x}^{t+1})=\bm{r}(\bm{x}^t)$ of period $n \geq 2$ is a ``variety'' of periodic points 
$$
\Set{ \bm{x} \in \mathbb{C}^d | F^{(n)}(\bm{x})-\bm{x}=0, \ F^{(m)}(\bm{x})-\bm{x} \neq 0, \ m \leq n },
$$
which is given by only the invariants
\begin{equation}
\Set{ \bm{x} \in \mathbb{C}^d | \bm{\gamma}^{(n)}(\bm{r}(\bm{x}))=0 }, \quad \bm{\gamma^{(n)}} \circ \bm{r} : \mathbb{C}^d \rightarrow \mathbb{C}^{d-p}.
\label{IVPP}
\end{equation}
The IVPP has an important property which is called the IVPP theorem\cite{IVPPa}\cite{IVPPb}\cite{IVPPbook}:

\bigskip

{\it \noindent Let F be a $d$ dimensional ``rational'' map with $p$ invariants. If $p \geq d/2$, an IVPP and discrete periodic points on a level set of any period do not exist in one map, simultaneously.}

\bigskip

\noindent The IVPP theorem gives an essential information of the integrability of the map. Because a Julia set\cite{Chaos} which characterizes a non integrable system is given by the closure of the sum set of discrete ``repulsive'' periodic points on a level set.

Therefore it seems that the existence of IVPP/Julia set is incompatible with the existence of Julia set/IVPP. This conjecture leads to the sufficient conditions of the integrability of a map as follows:

\bigskip

{\it \noindent If a map $F$ has an IVPP/Julia set then the map $F$ is Integrable/Non Integrable.}

\bigskip

In particular, the reason why the IVPP is interesting, is a phenomenon that periodic points become a set of continuum points.

There have been derived many IVPPs of various maps in the form of algebraic varieties, however, there have never been investigated the structure of an IVPP in detail. One of our purposes of this paper is to explore the component structure of an IVPP.
On the other hand it was known\cite{RE2} that, when a map is restricted on the IVPP of period $n$ it provides a recurrence equation\cite{RE0}\cite{RE1} of period $n$, an equation whose solutions are $n$ periodic for all initial points.

We show that a $d$ dimensional recurrence equation has a discrete symmetry, hence it gives a decomposition of the $d$ dimensional space to ``fundamental domains''.
In other words, components of IVPP are given by fundamental domains of the recurrence equation which is a restriction map on IVPP.
Hence comparing these two objects we will find a relation of components of IVPP and fundamental domains of recurrence equation as follows:

\bigskip

\begin{center}
{\it Components of IVPP $\sim$ Fundamental domains of recurrence equation}
\end{center}

\bigskip

In addition, we give an algorithm to derive components of IVPP for all rational maps at the final section.

This structure tells us an important information, how points move on IVPPs. In this paper, we show two concrete examples of such component structure of IVPPs which are restricted by the cyclotomic polynomial\cite{cp} of the recurrence maps.

\section{2 dimensional Case}
First, we introduce a 2 dimensional map and its IVPPs\cite{sshyw}.

\subsection{Map}
\begin{equation}
F_{2d} : (x, y) \mapsto (X, Y):=\left( x\frac{1-y}{1-x}, y\frac{1-x}{1-y} \right), \quad x, y \in \mathbb{C},
\label{map}
\end{equation}
with an invariant $r:=r(x,y)$
$$
r=xy.
$$

\subsection{IVPPs}
We know general formula of the IVPPs(\ref{IVPP}) of the map(\ref{map}) as follows\cite{sshyw}:
\begin{equation}
\gamma^{(n)}(r) =r+ \tan^2 \left( \frac{\pi m}{n} \right), \quad m= 1,2 \dots, n-1, \quad n=3,4, \dots,
\end{equation}
and explicit forms, which will be discussed in this paper, as follows:
\begin{eqnarray}
\gamma^{(2)}(r) &=& none, \nonumber \\
\gamma^{(3)}(r) &=& 3+r, \nonumber \\
\gamma^{(4)}(r) &=& 1+r, \nonumber \\
\gamma^{(5)}(r) &=& 5+10r+r^2, \nonumber \\
\gamma^{(6)}(r) &=& 1+3r, \nonumber \\
\mathrm{etc...} \nonumber
\end{eqnarray}
\begin{center}
\begin{figure}[H]
\begin{center}
\includegraphics[scale=0.35, bb=0 0 549 563]{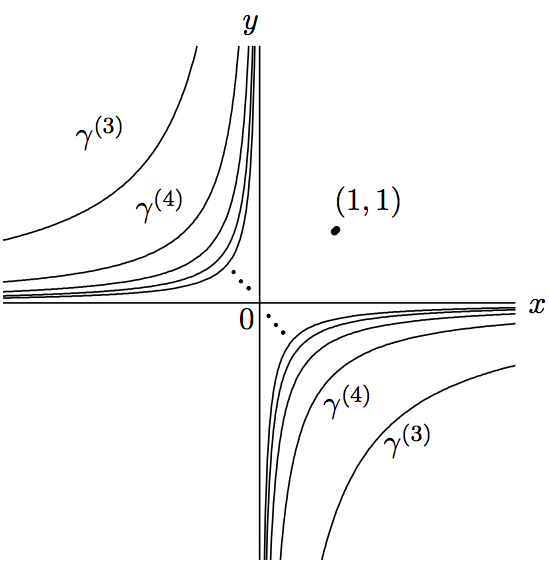}
\caption{IVPPs of the map (\ref{map})}
\end{center}
\end{figure}
\end{center}

\subsection{Components of IVPPs}
In this subsection, we give components of the IVPP of the map(\ref{map}).

\subsubsection{IVPP of period 3}
First, we give the flow of points on the IVPP of period 3 by symbolic calculus which is parametrized by $x$ as follows:
\begin{equation}
\left(x, -\frac{3}{x} \right) \rightarrow \left( \frac{3+x}{1-x}, 3\frac{1-x}{3+x} \right) \rightarrow \left( \frac{x-3}{1+x}, 3\frac{1+x}{x-3} \right) \rightarrow \left(x, -\frac{3}{x} \right).
\label{seq}
\end{equation}
Now, we can give boundaries of components of the IVPP of period 3 in the $x$ direction by the substitution $-\infty$ for $x$ of (\ref{seq}). Therefore we get the components of the IVPP of period 3 in the $x$ direction $(C_{3i})_x, i=1,2,3$ as follows,
$$
(C_{31})_x =(-\infty, -1), \quad (C_{32})_x = [-1, 1), \quad (C_{33})_x = [1, \infty),
$$
$$
C_{31} \overset{F}{\longrightarrow} C_{32} \overset{F}{\longrightarrow} C_{33} \overset{F}{\longrightarrow} C_{31}.
$$
We can draw a picture of ``tiling'' of the IVPP of period 3.
\begin{center}
\begin{figure}[H]
\begin{center}
\includegraphics[scale=0.6, bb=0 0 416 379]{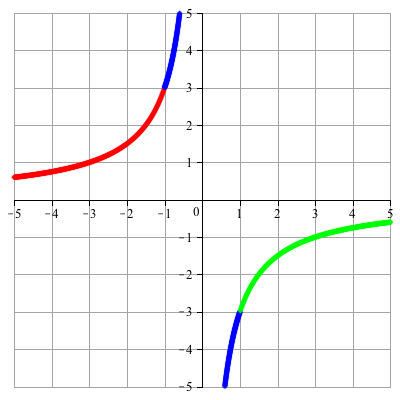}
\caption{Components of IVPP of period 3, $C_{31}$ : Red, $C_{32}$ : Blue, $C_{33}$ : Green.
}
\end{center}
\end{figure}
\end{center}

\subsubsection{IVPP of period 4}
All other periods are to be discussed in almost the same way of the case of period 3:

\bigskip

\noindent Flow of points on the IVPP of period 4:
\begin{eqnarray}
\left(x, -\frac{1}{x} \right) &\rightarrow& \left( \frac{1+x}{1-x}, -\frac{1-x}{1+x} \right) \rightarrow \left( -\frac{1}{x}, x \right) \nonumber \\
&\rightarrow& \left( -\frac{1-x}{1+x}, \frac{1+x}{1-x} \right) \rightarrow \left(x, -\frac{1}{x} \right). \nonumber
\end{eqnarray}

\noindent Components of the IVPP of period 4 in the $x$ direction $(C_{4i})_x, i=1,\dots, 4$ as follows:
$$
(C_{41})_x:= (-\infty, -1), \quad (C_{42})_x:= [-1, 0), \quad (C_{43})_x:= [0, 1), \quad (C_{44})_x:= [1, \infty),
$$
$$
C_{41} \overset{F}{\longrightarrow} C_{42} \overset{F}{\longrightarrow} C_{43} \overset{F}{\longrightarrow} C_{44} \overset{F}{\longrightarrow} C_{41}.  
$$
\begin{center}
\begin{figure}[H]
\begin{center}
\includegraphics[scale=0.6,bb=0 0 416 379]{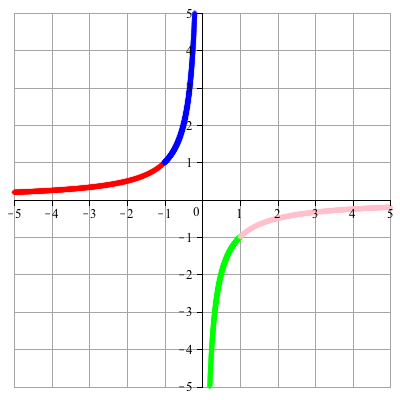}
\caption{Components of IVPP of period 4, $C_{41}$ : Red, $C_{42}$ : Blue, $C_{43}$ : Green, $C_{44}$ : Pink.}
\end{center}
\end{figure}
\end{center}

\subsubsection{IVPP of period 5}
The flow of points on the IVPP of period 5 are as follows,
\begin{eqnarray}
\left(x, \frac{a_{\pm}}{x} \right) &\rightarrow& \left( \frac{-a_{\pm}+x}{1-x}, \frac{a_{\pm}(1-x)}{-a_{\pm}+x} \right) \rightarrow \left( \frac{b_{\pm}(x\pm \sqrt{5})}{b_{\pm}-x}, \frac{a_{\pm}(b_{\pm}-x)}{b_{\pm}(x\pm \sqrt{5})} \right) \nonumber \\
&\rightarrow& \left( \frac{b_\pm (x \mp \sqrt{5})}{b_\pm+x }, -\frac{a_\pm(b_\pm+x) }{b_\pm (x \mp \sqrt{5})} \right) \rightarrow \left( \frac{a_\pm+ x}{1+x}, \frac{a_\pm (1+x)}{a_\pm +x} \right) \nonumber \\
&\rightarrow & \left(x, \frac{a_{\pm}}{x} \right), \nonumber
\end{eqnarray}
where $a_{\pm}:=-5\pm 2\sqrt{5}, \ b_{\pm}:=-2\pm \sqrt{5}$. We notice that the IVPP of period 5 is multivalued dependent on $a_{\pm}$, hence we must discuss them separately.

We get components of IVPP of period 5 $(C_{5_{\pm} i})_x, i=1,\dots, 5$ as follows,
\begin{eqnarray}
\mathrm{A \ part \ of \ } a_+ \ &:&  \ (C_{5_+ 1})_x:=(-\infty, -1), \quad (C_{5_+ 2})_x:=[-1, -b_+), \quad (C_{5_+ 3})_x:=[-b_+, b_+), \nonumber \\
&& \ (C_{5_+ 4})_x:=[b_+, 1), \quad (C_{5_+ 5})_x:=[1, \infty), \nonumber \\
\mathrm{A \ part \ of \ } a_- \ &:&  \ (C_{5_- 1})_x:=(-\infty, b_-), \quad (C_{5_- 2})_x:=[b_-, -1), \quad (C_{5_- 3})_x:=[-1, 1), \nonumber \\
&& \ (C_{5_- 4})_x:=[1, -b_-), \quad (C_{5_- 5})_x:=[-b_-, \infty), \nonumber
\end{eqnarray}
\begin{eqnarray}
\mathrm{A \ part \ of \ } a_+ \ &:& \ C_{5_+ 1} \overset{F}{\longrightarrow} C_{5_+  2} \overset{F}{\longrightarrow} C_{5_+  3} \overset{F}{\longrightarrow} C_{5_+ 4} \overset{F}{\longrightarrow} C_{5_+ 5} \overset{F}{\longrightarrow} C_{5_+ 1}, \nonumber \\
\mathrm{A \ part \ of \ } a_- \ &:& \ C_{5_-  1} \overset{F}{\longrightarrow} C_{5_- 3} \overset{F}{\longrightarrow} C_{5_- 5} \overset{F}{\longrightarrow} C_{5_ - 2} \overset{F}{\longrightarrow} C_{5_-  4} \overset{F}{\longrightarrow} C_{5_- 1}. \nonumber
\end{eqnarray}
\begin{center}
\begin{figure}[H]
\begin{center}
\includegraphics[scale=0.6,bb=0 0 416 379]{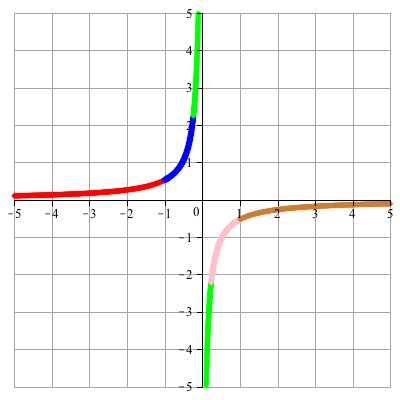}
\caption{A part of $a_+$ of components of IVPP of period 5, $C_{5_{+}1}$ : Red, $C_{5_{+}2}$ : Blue, $C_{5_{+}3}$ : Green, $C_{5_{+}4}$ : Pink, $C_{5_{+}5}$ : Gold.}
\end{center}
\end{figure}
\end{center}
\begin{center}
\begin{figure}[H]
\begin{center}
\includegraphics[scale=0.6,bb=0 0 416 379]{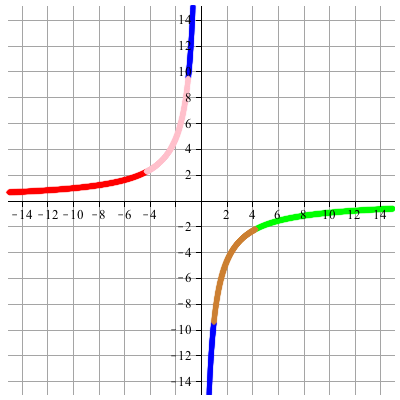}
\caption{A part of $a_-$ of components of IVPP of period 5, $C_{5_{-}1}$ : Red, $C_{5_{-}2}$ : Blue, $C_{5_{-}3}$ : Green, $C_{5_{-}4}$ : Pink, $C_{5_{-}5}$ : Gold.}
\end{center}
\end{figure}
\end{center}

\subsubsection{IVPP of period 6}
The flow of points on the IVPP of period 6 is as follows:
\begin{eqnarray}
\left(x, -\frac{1}{3x} \right) &\rightarrow& \left( \frac{1+3x}{3(1-x)}, -\frac{1-x}{1+3x} \right) \rightarrow \left( \frac{1+x}{1-3x}, -\frac{1-3x}{1+x} \right) \nonumber \\
&\rightarrow& \left( -\frac{1}{3x}, x \right) \rightarrow \left( -\frac{1-x}{1+3x}, \frac{1+3x}{3(1-x)} \right) \nonumber \\
&\rightarrow & \left( -\frac{1-3x}{1+x}, \frac{1+x}{1-3x} \right) \rightarrow \left(x, -\frac{1}{3x} \right). \nonumber
\end{eqnarray}
Therefore we get components of IVPP of period 6 as follows:
\begin{eqnarray}
&& (C_{61})_x:=(-\infty, -1), \quad (C_{62})_x:=[-1, -1/3), \quad (C_{63})_x:=[-1/3, 0), \nonumber \\
&& (C_{64})_x:=[0, 1/3), \quad (C_{65})_x:=[1/3, 1), \quad (C_{66})_x:=[1, \infty), \nonumber
\end{eqnarray}
$$
C_{61} \overset{F}{\longrightarrow} C_{62} \overset{F}{\longrightarrow} C_{63} \overset{F}{\longrightarrow} C_{64} \overset{F}{\longrightarrow} C_{65} \overset{F}{\longrightarrow} C_{66} \overset{F}{\longrightarrow} C_{61}.
$$
\begin{center}
\begin{figure}[htbp]
\begin{center}
\includegraphics[scale=0.6,bb=0 0 416 379]{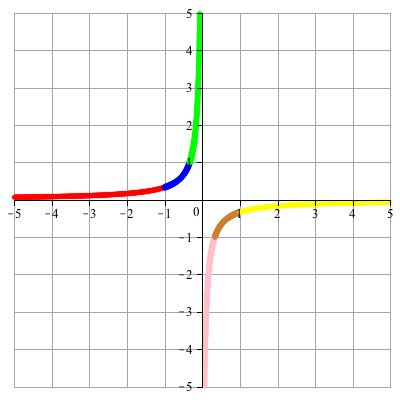}
\caption{Components of IVPP of period 6, $C_{61}$ : Red, $C_{62}$ : Blue, $C_{63}$ : Green, $C_{64}$ : Pink, $C_{65}$ : Gold, $C_{66}$ : Yellow.}
\end{center}
\end{figure}
\end{center}

\section{The component structure of IVPPs of the map (\ref{map})}
In this section, we discuss the structure of the components of IVPPs of the map (\ref{map}). The map(\ref{map}) is reduced by the invariant as follows:
$$
x \mapsto X=\frac{x-r}{1-x}, \quad x, X \in \mathbb{C},
$$
or in the linear map form
\begin{equation}
\left( 
\begin{array}{cc} 
X \\ 1\\ 
\end{array} \right) 
=
M
\left( 
\begin{array}{cc} 
x \\
1 \\
\end{array}
\right), \quad \left( 
\begin{array}{cc} 
x \\ 1\\ 
\end{array} \right),
\left( 
\begin{array}{cc} 
X \\ 1\\ 
\end{array} \right)
\in \mathbb{C}\mathbb{P}^2,
\label{mob}
\end{equation}
where $M=\left(
\begin{array}{cc} 
1 & -r \\ 
-1 & 1 \\ 
\end{array}
\right) $.
It is a M\"{o}bius map. Furthermore, we can transform $M$ to a diagonal matrix by coordinate transformation as
$$
\left( 
\begin{array}{cc} 
z \\ 1\\ 
\end{array} \right) 
=
\left(
\begin{array}{cc} 
-\sqrt{r} & \sqrt{r} \\ 
1 & 1 \\ 
\end{array}
\right) 
\left( 
\begin{array}{cc} 
x \\
1 \\
\end{array}
\right), \quad \left( 
\begin{array}{cc} 
x \\ 1\\ 
\end{array} \right),
\left( 
\begin{array}{cc} 
z \\ 1\\ 
\end{array} \right)
\in \mathbb{C}\mathbb{P}^2.
$$
Therefore, we get a diagonal form of (\ref{mob}) as follows:
\begin{equation}
\left( 
\begin{array}{cc} 
Z \\ 1\\ 
\end{array} \right) 
=
\left(
\begin{array}{cc} 
\lambda_+ & 0 \\ 
0 & \lambda_- \\ 
\end{array}
\right) 
\left( 
\begin{array}{cc} 
z \\
1 \\
\end{array}
\right), \quad \left( 
\begin{array}{cc} 
z \\ 1\\ 
\end{array} \right),
\left( 
\begin{array}{cc} 
Z \\ 1\\ 
\end{array} \right)
\in \mathbb{C}\mathbb{P}^2,
\label{mob2}
\end{equation}
where the $\lambda_{\pm}:=1\pm \sqrt{r}$ are eigenvalues of $M$.
In other words, if we take a new invariant
\begin{equation}
s=\frac{\lambda_+}{\lambda_-}, \quad \lambda_{\pm} \in \mathbb{C},
\label{newinv}
\end{equation}
then the map (\ref{mob2}) is just a scale transformation
$$
z \mapsto Z = sz, \quad z, Z \in \mathbb{C}.
$$
This fact means that the IVPP of (\ref{mob2}) of period $n$ is given by a cyclotomic polynomial\cite{cp}:
$$
\Phi_n(s) \sim s^n-1=0.
$$
Hence the components of the IVPP of period $n$ of the map(\ref{mob2}) is equivalent to an orbifold $\mathbb{C}/\mathbb{Z}_n$:
\begin{figure}[H]
\begin{tabular}{cc}
\begin{minipage}{0.5\hsize}
\begin{center}
\includegraphics[scale=0.44, bb=0 0 416 379]{decIVPP3.png}
\caption{IVPP of period 3 on real space}
\end{center}
\end{minipage}
\begin{minipage}{0.5\hsize}
\begin{center}
\includegraphics[scale=0.44, bb=0 0 374 415]{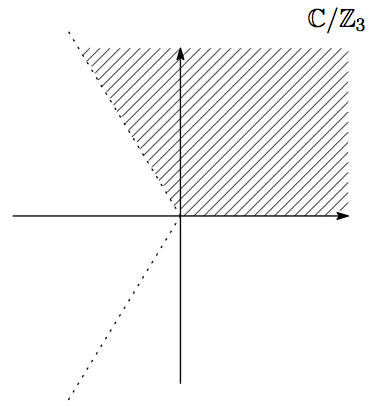}
\caption{$\mathbb{C}/\mathbb{Z}_3$}
\end{center}
\end{minipage}
\end{tabular}
\end{figure}

We give a formula of $x$ direction boundaries $c_m$ of the components of IVPPs of period $n$ as follows:
\begin{equation}
c_m:= \frac{(1-s_n)(1+s_n^m)}{(1+s_n)(1-s_n^m)}, \quad m=0,1, \dots, n,
\label{formula}
\end{equation}
where $s_n$ satisfies the $n$-order cyclotomic polynomial. Furthermore, a formula in $z$ space is also given by
\begin{equation}
d_m := -\sqrt{r} \frac{\sqrt{r}(\lambda_+^m+\lambda_-^m)+(\lambda_+^m-\lambda_-^m)}{\sqrt{r}(\lambda_+^m+\lambda_-^m)-(\lambda_+^m-\lambda_-^m)}, \quad m=0,1, \dots, n.
\label{formula2}
\end{equation}
These formulas are given by the pull buck the $m$ iterated map (\ref{mob2}) to the $m$ iterated map (\ref{mob}) as follows,
\begin{eqnarray}
\left(
\begin{array}{cc} 
1 & -r \\ 
-1 & 1 \\ 
\end{array}
\right)^m 
&=& 
O \left(
\begin{array}{cc} 
\lambda_+^m & 0 \\ 
0 & \lambda_-^m \\ 
\end{array}
\right) 
O^{-1} \nonumber \\
&=& \left(
\begin{array}{cc} 
\lambda_+^m+\lambda_-^m & -\sqrt{r}(\lambda_+^m-\lambda_-^m) \\ 
-\frac{1}{\sqrt{r}}(\lambda_+^m-\lambda_-^m) & \lambda_+^m+\lambda_-^m \\ 
\end{array}
\right), \nonumber
\end{eqnarray}
where $O$ is the diagonalization matrix of $M$. The $x$ direction boundaries are given by the substitution $-\infty$ for $x$. Therefore we can get the formula(\ref{formula}).

Finally, we give one more point of view about the boundaries of components of the IVPP. We substitute $-\infty$ for $x$ of the flow on the IVPP, hence the boundaries are given by the ``intersections of the IVPP and zero points set of denominators of iterations of the map''. We draw a graph of zero points set of the denominators of iterations of the map (\ref{map}) as follows:
\begin{center}
\begin{figure}[H]
\begin{center}
\includegraphics[scale=0.6,bb=0 0 400 400]{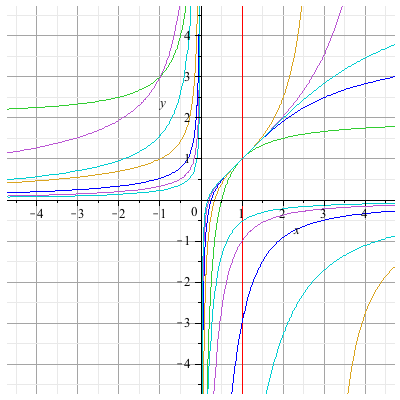}
\caption{Zero points set of denominators of iterations of the map (\ref{map}), $F$ : Red, $F^{(2)}$ : Green, $F^{(3)}$ : Yellow, $F^{(4)}$ : Blue, $F^{(5)}$ : Purple, $F^{(6)}$ : Light blue.}
\end{center}
\end{figure}
\end{center}
%

\section{3 Dimensional Case}
Next, we try to discuss 3 dimensional case. In particular, we discuss about the 3 dimensional Lotka-Volterra map\cite{LV}, and only the IVPP of period 2, because higher dimensional cases and higher period cases are difficult to write symbolic calculus and decompositions.

\subsection{Map}
\begin{equation}
F_{3d} : (x,y,z) \to (X,Y,Z)=\left(x{1-y+yz\over 1-z+zx},\ y{1-z+zx\over 1-x+xy},\ z{1-x+xy\over 1-y+yz}\right), \quad x,y,z \in \mathbb{C},
\label{LVmap}
\end{equation}
with invariants $r:=r(x,y,z), s:=s(x,y,z)$,
\begin{eqnarray*}
r&=&xyz, \nonumber \\
s&=&(1-x)(1-y)(1-z).
\end{eqnarray*}

\subsection{IVPPs}
\begin{center}
\begin{figure}[H]
\begin{center}
\includegraphics[scale=0.5,bb=0 0 400 400]{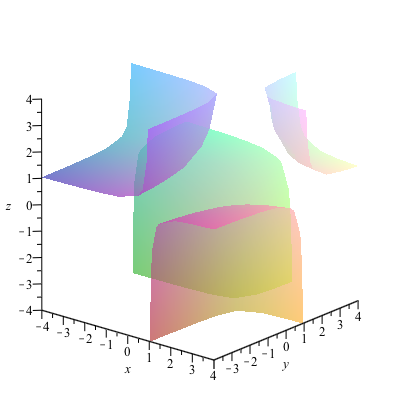}
\caption{IVPP of period 2}
\label{LVIVPP}
\end{center}
\end{figure}
\end{center}

We know some conditions of the IVPPs of the map (\ref{LVmap}) as follows\cite{IVPPb}:
\begin{eqnarray*}
\gamma^{(2)}(r, s) &=&s+1,\\
\gamma^{(3)}(r, s) &=&(s-r)^2+(r+1)(s+1),\\
\gamma^{(4)}(r, s) &=&(s-r)^3+s(r+1)^3,\\
{\rm etc.},&&
\end{eqnarray*}
We draw the IVPP of period 2 in FIG. \ref{LVIVPP}.
%

\subsection{Components of IVPP of Period 2}
Similar to the case of 2 dimensional map, we give a symbolic calculus on IVPP of period 2. 
\begin{eqnarray}
\left( x, \frac{a_{\pm}}{x-1}, \frac{a_{\mp}}{x-1} \right) \rightarrow \left( \frac{x}{x-1}, a_{\mp}, a_{\pm} \right),
\end{eqnarray}
where 
$$
a_{\pm}= \frac{-r+rx+2x-x^2 \pm \sqrt{r^2-2r^2x+2rx^2+r^2x^2-2rx^3+4x^2-4x^3+x^4}}{2x}.
$$
This case also is multivalued. Therefore we get components of IVPP of period 2 as follows:
$$
[(C_{2_\pm 1})_1]_x= (-\infty, 0], \quad [(C_{2_\pm 2})_1]_x= (0, 1], \quad [(C_{2_\pm 1})_2]_x= (1, \infty], \quad [(C_{2_\pm 2})_2]_x= (1, \infty].
$$
\begin{center}
\begin{figure}[H]
\begin{center}
\includegraphics[scale=0.5,bb=0 0 400 400]{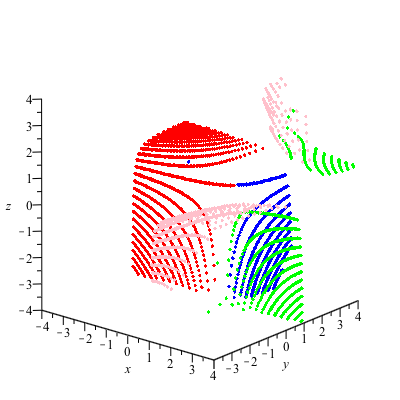}
\caption{A part of $a_+$ of components of IVPP of period 2, $(C_{2_+1})_1$: Red, $(C_{2_+2})_2$: Blue, $(C_{2_+1})_1$: Green, $(C_{2_+1})_2$: Pink.}
\end{center}
\end{figure}
\end{center}
\begin{center}
\begin{figure}[H]
\begin{center}
\includegraphics[scale=0.5,bb=0 0 400 400]{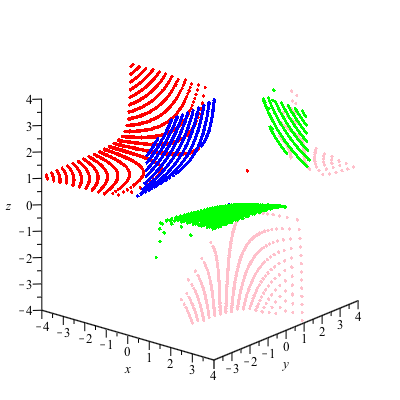}
\caption{A part of $a_-$ of components of IVPP of period 2, $(C_{2_-1})_1$: Red, $(C_{2_-2})_2$: Blue, $(C_{2_-1})_1$: Green, $(C_{2_-1})_2$: Pink.}
\end{center}
\end{figure}
\end{center}
Here, we notice that each graph has two kinds of tiling, {\it i.e.,} red and blue pair and green pink pair in both the FIG. 11 and FIG. 12. It is a new phenomenon.

In addition, we should give a remark about the reason of appearance of the striped pattern of tilings of the IVPP. The IVPP of period 2 is given by the condition as follows,
$$
s+1=0 \quad \Rightarrow \quad s=-1,
$$
then the IVPP of period 2 has a free parameter $r$ which is not restricted by this condition. Therefore we draw graphs by showing stripes corresponding to the parameter of invariant $r$ which is stepped by $1$ for visibility. In other words, each stripe is one of a level set corresponding to a part of IVPP. In general, any IVPPs of the map(\ref{LVmap}) are given by one condition, hence we can solve the condition about $s$. In other words, any IVPPs are described by one parameter $r$.

\subsection{Remarks}
We give a small remark about the relation of decompositions of IVPP and fundamental domains of the recurrence equation. The recurrence equation of period 2 is a restriction on IVPP of period 2 of the map(\ref{LVmap}) as follows:
\begin{equation}
x \mapsto X= -\frac{x}{1-x}, \quad x, X \in \mathbb{C}.
\label{LVre}
\end{equation}
We notice that the map(\ref{LVre}) is independent on the parameter $r$. In a similar way as the case of 2 dimensional map(\ref{map}), we can get a diagonal form
$$
w \mapsto W= -w, \quad w, W \in \mathbb{C},
$$
and the fundamental domain is given by $\mathbb{C}/ \mathbb{Z}_2$.

\section{Conclusion}

In order to conclude this paper we would like to give our algorithm to determine the components of IVPP for all rational maps:

\noindent {\it Algorithm}
\begin{itemize}
\item[1] Give an initial point $\bm{x}^0$ on IVPP. If IVPP is multivalued, then we must use our algorithm for all parts.
\item[2] Symbolic calculation of the mapping on the IVPP of period $n$:
$$
\bm{x}^0 \rightarrow \bm{x}^1(\bm{x}^0) \rightarrow \dots \rightarrow \bm{x}^{n-1}(\bm{x}^0) \rightarrow \bm{x}^n(\bm{x}^0)=\bm{x}^0.
$$
\item[3] Substitution for a minimum value of domain of parameters $\tilde{\bm{x}} \in \mathbb{C}^{d-p}, \tilde{\bm{r}} \in \mathbb{C}^{2p-d}$ of the IVPP of period $n$ for any points $\bm{x}^t, t=0, 1, \dots, n$.
\item[4] Changing the parameters of step 2 to positive direction, until go to other components.
\item[5] If there exists a non discussed domain when finished the above operations, then go to step 3 for non discussed domain.
\item[6] Therefore we get component family of the IVPP of period $n$ in the $\tilde{\bm{x}}$ direction about parameters $\tilde{\bm{r}}$.
\end{itemize}

In conclusion, above discussions give a relation of decomposition of IVPP and the tiling as follows:
$$
\mathrm{IVPP \ of \ period} \ n \sim \mathbb{C}^{d-p}/ \mathbb{Z}_n \times \mathbb{C}^{2p-d} \times \mathrm{number \ of \ tiling }.
$$

\section*{Acknowledgement}
I would like to thank Dr. S. Saito and Mr. Y. Wakimoto for useful discussion.


\appendix
\section{Reason of missing the IVPP of period 2}
We do not discuss a reason of missing the IVPP of the map(\ref{mob}) of period 2. This reason is given from the transformation of invariant(\ref{newinv})
$$
s=\frac{1+\sqrt{r}}{1-\sqrt{r}}.
$$
By the cyclotomic polynomial, the IVPP of the map(\ref{mob2}) of period 2 is given $s=-1$. Therefore if $s=-1$ then $r$ is satisfied $r=\infty$ by (\ref{newinv}).

\section{A relation of $x$ and $z$}
We check the values of $z$ when $x$ is (real) boundary of components of IVPPs. These are given by the formula (\ref{formula2}):
\begin{itemize}
\item Period 3:
$$
x=\pm \infty \Rightarrow z=-\sqrt{3}i, \quad x=-1 \Rightarrow z=\infty, \quad x=1 \Rightarrow z=0,
$$

\item Period 4:
$$
x=\pm \infty \Rightarrow z=-i, \quad x=-1 \Rightarrow z=\infty, \quad x=0 \Rightarrow z=i \quad x=1 \Rightarrow z=0,
$$

\item Period 5:
	\begin{itemize}
	\item A part of $a_+$ :
	\begin{eqnarray}
&&x=\pm \infty \Rightarrow z=-\sqrt{a_+}, \quad x=-1 \Rightarrow z=\infty, \quad x=-b_+ \Rightarrow z=\frac{1}{2}\sqrt{a_+}(\sqrt{5}+1), \nonumber \\
&&x=b_+ \Rightarrow z=\frac{1}{2}\sqrt{a_+}(\sqrt{5}-1),  \quad x=1 \Rightarrow z=0, \nonumber
\end{eqnarray}
	\item A part of $a_-$ :
	\begin{eqnarray}
&&x=\pm \infty \Rightarrow z=-\sqrt{a_-}, \quad x=-1 \Rightarrow z=\infty, \quad x=-b_- \Rightarrow z=-\frac{1}{2}\sqrt{a_-}(\sqrt{5}-1), \nonumber \\
&&x=b_- \Rightarrow z=-\frac{1}{2}\sqrt{a_-}(\sqrt{5}+1),  \quad x=1 \Rightarrow z=0, \nonumber
\end{eqnarray}
	\end{itemize}
\item Period 6:
\begin{eqnarray}
&&x=\pm \infty \Rightarrow z=-\frac{\sqrt{3}}{3}i, \quad x=-1 \Rightarrow z=\infty, \quad x=-\frac{1}{3} \Rightarrow z=\frac{2\sqrt{3} }{3} i, \nonumber \\
&& x=0 \Rightarrow z=\frac{\sqrt{3} }{3} i, \quad x=\frac{1}{3} \Rightarrow z=\frac{\sqrt{3} }{6} i,  \quad x=1 \Rightarrow z=0. \nonumber
\end{eqnarray}
\end{itemize}


\begin{thebibliography}{00}
\bibitem{IVPPa}
S. Saito and N. Saitoh, {\it J. Phys. Soc. Jpn.}, {\bf 76} No.2 p.024006, 2007.

\bibitem{IVPPb}
S. Saito and N. Saitoh {\it J. Math. Phys}, {\bf 51} 063501, 2010.

\bibitem{IVPPbook}
S. Saito and N. Saitoh, {\it ``Invariant varieties of periodic points''} in
Mathematical Physics Research Developments, 2008 Nova Science Publishers, Inc., Capt.3 pp 85-139, 2008.

\bibitem{cp}
Wolfram Mathworld, mathworld.wolfram.com/CyclotomicPolynomial.html.

\bibitem{Chaos}
R. L. Devaney, {\it ``An Introduction to Chaotic Dynamical Systems''}, Westview Press, 2003.

\bibitem{RE0}
R. L. Graham, D. E. Knuth and O. Patashnik, {\it Concrete Mathematics} (Addison-Wesley), 1994.
 
\bibitem{RE1}
R. Hirota and H. Yahagi, {\it J. Phys. Soc. Jpn.}, {\bf 71}, 2867, 2002.
 
\bibitem{RE2}
S. Saito and N. Saitoh, {\it J. Phys. A: Math. Theor.}, {\bf 40}, 12775-12787, 2007.

\bibitem{sshyw}
S. Saito, N. Saitoh, H. Harada, T. Yumibayashi and Y. Wakimoto, AIP Advances, AIP ID: 003306ADV, 2013.

\bibitem{LV}
R. Hirota, S. Tsujimoto and T. Imai,  {\it Future Directions of Nonlinear Dynamics in Physical and Biological Systems}, ed. by P.L.Christiansen at al., p.7 (Plenum Press, New York, 1993). 

R. Hirota, and S. Tsujimoto,  {\it J.Phys.Soc.Jpn.} {\bf 64} 3125-3127 (1995).

\end{thebibliography}
\end{document}